\documentclass[11pt,epsfig]{article}
\usepackage{graphicx}
\begin{document}
\title{\bf Anomalous Magnetic Moment of Electron in
  Chern-Simons QED} 
\author{Ashok Das$^{\rm a}$ and Silvana Perez$^{\rm b}$\\
\\
$^{\rm a}$ Department of Physics and Astronomy,\\
University of Rochester,\\
Rochester, NY 14627-0171,\\
USA\\
\\
$^{\rm b}$ Departamento de F\'{\i}sica,\\
Universidade Federal do Par\'{a},\\
Bel\'{e}m, PA 66075-110,\\
BRAZIL}

\date{}
\maketitle

\begin{center}
{ \bf Abstract}
\end{center}

We calculate the anomalous magnetic moment of the
electron in the Chern-Simons theory in $2+1$ dimensions with and without a
Maxwell term, both at
zero temperature as well as at finite temperature. In the case of
the Maxwell-Chern-Simons (MCS) theory, we find that there is an infrared
divergence, both at zero as well as at
finite temperature, when the tree level Chern-Simons term
vanishes, which suggests that a Chern-Simons term is essential in such
theories.  At high temperature, the thermal correction in the MCS
theory behaves as $\frac{1}{\beta} \ln \beta M$, where $\beta$ denotes
the inverse temperature and $M$, the Chern-Simons coefficient. On the
other hand, we find no thermal correction to the anomalous magnetic
moment in the pure Chern-Simons (CS) theory.

\newpage

\section{Introduction}

Chern-Simons theories \cite{deser,hagen} in $2+1$ dimensions have been
of interest from a
variety of reasons and have led to many interesting phenomena in
various branches of physics \cite{dunne}. In this paper, we
systematically study
the question of the anomalous magnetic moment of an electron in the
Chern-Simons theory with and without a Maxwell term, both at zero as
well as at finite temperature. In an Abelian theory, the magnetic moment is a
gauge invariant quantity much like the Chern-Simons coefficient. As a
result, its value is not related by the Ward identity to any other
amplitude in the theory and needs to be studied independently. Our
study brings out some new and interesting properties of Chern-Simons
theories. For example, although
conventionally one argues that the topological Chern-Simons term is an
additional gauge invariant term that can be added to the action, the
study of the anomalous magnetic moment, both at zero as well as finite
temperature, reveals that such a term is, in fact, essential and
in the absence of this term, physical quantities such as the
magnetic moment cannot be defined because of infrared
divergence. This has to be contrasted with the magnetic moment in
$3+1$ dimensions where it is both ultraviolet and infrared
finite. Normally, one expects infrared divergence to be more
severe in lower dimensions. However, in $2+1$ dimensions it is
believed that at least the Abelian theory has a
well defined infrared behavior \cite{jackiw}, unless one is
at high temperature \cite{dunne1}  and at higher loops \cite{brandt1}.
Non-Abelian
theories, of course, can have stronger infrared divergent
behavior \cite{brandt}. Our result, on the other hand
shows that the Abelian theory can also exhibit infrared divergence at
zero temperature at one loop in
physical quantities, such as the anomalous magnetic moment unless
there is a tree level Chern-Simons mass.

In $3+1$ dimensional QED, anomalous magnetic moment at zero
temperature  is one of the most intensively studied quantities both
theoretically and experimentally where the value of $(g-2)$ is quite
small \cite{itzykson}. In contrast, we find that in $2+1$ dimensions,
the value of
$(g-2)$ can become arbitrarily large depending on the value of the
Chern-Simons coefficient. The thermal behavior of the anomalous
magnetic moment in $3+1$ dimensions has also been studied
\cite{yee,barton} and the
temperature dependence determined to be $\frac{1}{\beta^{2}}$ at high
temperatures, where $\beta$ denotes the inverse temperature. In $2+1$
dimensional Maxwell-Chern-Simons theory, we find that the thermal
correction to the anomalous magnetic moment goes as $\frac{1}{\beta}
\ln \beta M$ where $M$ represents the Chern-Simons
coefficient. Furthermore, in the pure Chern-Simons theory, the
anomalous magnetic moment surprisingly has no thermal correction.

The anomalous magnetic moment in $2+1$ dimensions has been studied
earlier at zero temperature at one loop and the explicit value
obtained in the 
large $M$ limit (anyonic limit or the pure CS limit) using the
covariant Landau gauge \cite{kogan}. Subsequently, this result has
also been  derived 
from a calculation in the pure Chern-Simons theory in the Coulomb
gauge \cite{girotti}. Our study, on the other hand, represents a
complete systematic analysis of  the
anomalous magnetic moment at one loop in Chern-Simons theories with
and without the
Maxwell term, both at zero as well as at finite temperature. Our
presentation is organized as follows. In section {\bf 2}, we analyze
the question of the anomalous magnetic moment in the MCS theory at
zero temperature for an
arbitrary value of the Chern-Simons coefficient which brings out the
problem of the infrared divergence. Taking the large $M$
limit, we then obtain the known result for the pure CS theory
\cite{kogan,girotti}. In
section {\bf 3}, we present our analysis of the finite temperature
behavior of the anomalous magnetic moment. Here, the analysis for the
MCS theory and the pure CS theory need to be done separately as we
explain in the text. While the anomalous magnetic moment in the MCS
theory behaves as $\frac{1}{\beta} \ln \beta M$ at high temperature,
surprisingly  there is
no thermal correction to the anomalous magnetic moment in the pure CS
theory. We speculate on the reason for such a behavior and conclude with
a brief summary in section {\bf 4}. 

\section{Zero Temperature}

Let us consider the Maxwell-Chern-Simons theory described by the
Lagrangian density
\begin{equation}
{\cal L} = - \frac{1}{4} F_{\mu\nu}F^{\mu\nu} \pm \frac{M}{4}
\epsilon^{\mu\nu\lambda} A_{\mu} F_{\nu\lambda} + \overline{\psi}
(i\partial\!\!\!\slash - e A\!\!\!\slash - m)\psi - \frac{1}{2\xi}
(\partial_{\mu}A^{\mu})^{2}  +
 \partial^{\mu}\overline{c} \partial_{\mu} c\, .
\end{equation}
Here $\xi$ represents the gauge fixing parameter and we have allowed for both
signs of the Chern-Simons term. We assume that both the fermion mass
as well as the Chern-Simons coefficient $(m,M)$
are nonnegative and define for later convenience their ratio to be
\begin{equation}
\kappa = \frac{M}{m} \geq 0\, ,\label{ratio}
\end{equation}
which is a dimensionless constant. In the general covariant gauge, the
photon propagator is given by
\begin{equation}
iD_{\mu\nu} (p) = - \frac{i}{p^{2} - M^{2}}\left(\eta_{\mu\nu} -
  \frac{p_{\mu}p_{\nu}}{p^{2}} \pm iM
  \epsilon_{\mu\nu\lambda} \frac{p^{\lambda}}{p^{2}}\right) -
  i\xi\,\frac{p_{\mu}p_{\nu}}{(p^{2})^{2}}\, ,\label{photonpropagator}
\end{equation}
while the fermion propagator has the form (the $i\epsilon$
prescription is understood)
\begin{equation}
iS (p) = \frac{i}{p\!\!\!\slash - m} = \frac{i(p\!\!\!\slash +
  m)}{p^{2} - m^{2}}\, .
\end{equation}
We use the metric with signatures $(+,-,-)$ and a representation for
the gamma matrices with $\gamma^{0}$ Hermitian, $\gamma^{i}, i=1,2$
anti-Hermitian and satisfying
\begin{equation}
\gamma^{\mu}\gamma^{\nu} = \eta^{\mu\nu} - i \epsilon^{\mu\nu\lambda}
\gamma_{\lambda}\, .\label{gammaidentity}
\end{equation}

\begin{figure}[h]
\begin{center}
\includegraphics[width = 15cm, height = 8cm]{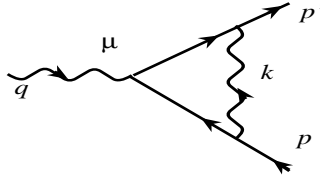}
\vspace{-2cm}
\caption{One-loop vertex correction.}
\end{center}
\end{figure}

The one loop correction to the vertex has the form
\begin{eqnarray}
\Gamma^{\mu} (p, p') & = & (-ie)^{3} \int
\mathrm{d}K\,\gamma^{\lambda} iS (p'-k)\gamma^{\mu} iS
(p-k)\gamma^{\rho} iD_{\rho\lambda} (k)\nonumber\\
\noalign{\vskip 4pt}%
 & = & e^{3} \int \mathrm{d}K\,\gamma^{\lambda}(p'\!\!\!\!\slash -
 k\!\!\!\slash + m)\gamma^{\mu} (p\!\!\!\slash - k\!\!\!\slash +
 m)\gamma^{\rho} D_{\rho\lambda} (k) \times \frac{1}{k^{2}-2k\cdot p}
 \frac{1}{k^{2}-2k\cdot p'}\, ,\label{vertex}
\end{eqnarray}
where $k$ is the internal loop momentum, $q$ the momentum transfer and
we are assuming that the external fermions are on-shell so that
\begin{equation}
q^{\mu} = p'^{\mu} - p^{\mu},\quad p^{2} = p'^{2} = m^{2},\quad {\rm
  and}\quad \mathrm{d}K = \frac{\mathrm{d}^{3}k}{(2\pi)^{3}}\, .
\end{equation}
At zero temperature, the vertex function can be parameterized as
\begin{equation}
\Gamma^{\mu} (p,p') = e \gamma^{\mu}\, F_{1} (q^{2}) +
\frac{ie}{4m}\,\left[\gamma^{\mu}, q\!\!\!\slash\right]\,F_{2}
(q^{2}) = e \gamma^{\mu}\, F_{1} (q^{2}) +
\frac{e}{2m}\,\sigma^{\mu\nu}q_{\nu}\, F_{2} (q^{2})\,
,\label{zeroTstructure}
\end{equation}
and the anomalous magnetic moment of the electron can be identified
with $F_{2} (q^{2}=0)$.

The photon propagator in (\ref{photonpropagator}) (and, therefore in
(\ref{vertex})) has a complex tensor
structure compared to usual QED. First, let us show that the
$p_{\mu}p_{\nu}$  terms in the
photon propagator contribute only to the charge
renormalization ($F_{1} (q^{2})$) and not to the magnetic moment. Replacing 
$D_{\rho\lambda} (k)\sim k_{\rho}k_{\lambda}$, the numerator in
(\ref{vertex}) takes the form
\begin{equation}
k\!\!\!\slash (p'\!\!\!\!\slash - k\!\!\!\slash + m) \gamma^{\mu}
(p\!\!\!\slash - k\!\!\!\slash + m) k\!\!\!\slash  = \left((k^{2})^{2}
- 2k\cdot (p'+p) k^{2} + 4k\cdot p' k\cdot p\right) \gamma^{\mu}\, ,
\end{equation}
which makes it clear that this can only contribute to the charge
renormalization ($F_{1} (q^{2})$) and not to the magnetic moment. In deriving
this, we
have used the fermion equations of motion at various intermediate
steps.  Thus, as far as the
magnetic moment calculation is concerned, we can equivalently, think
of the photon propagator as
\begin{equation}
D^{({\rm eff})}_{\mu\nu} (p) = - \frac{1}{p^{2} - M^{2}}
  \left(\eta_{\mu\nu} \pm iM
  \epsilon_{\mu\nu\lambda} \frac{p^{\lambda}}{p^{2}}\right)\,
  .\label{equivprop}
\end{equation}
This also makes it clear that the magnetic moment is manifestly 
independent of the gauge parameter, as one would expect for a physical
quantity. 

If we now look at the contribution coming only from the
  $\eta_{\rho\lambda}$ term in the photon propagator in
  (\ref{equivprop}), the numerator in (\ref{vertex}) takes the form
\begin{eqnarray}
N_{1}^{\mu} & = & \left(-4m^{2} - 2(q+k)^{2} + 3 k^{2}\right)
\gamma^{\mu}\nonumber\\
\noalign{\vskip 4pt}%
 &  & + 2\left[\gamma^{\mu}, q\!\!\!\slash\right] (-m + k\!\!\!\slash)
+ 4m (p'^{\mu}+p^{\mu}) - 4m k^{\mu} - 2 k^{\mu}k\!\!\!\slash\,
.\label{N1} 
\end{eqnarray}
Combining the denominators and shifting variables of integration in
(\ref{vertex}), we obtain
\begin{equation}
\Gamma_{(\eta)}^{\mu} = -2e^{3} \int_{0}^{1} \mathrm{d}x \int_{0}^{1-x}
\mathrm{d}y \int \mathrm{d}K\,\frac{\tilde{N}_{1}^{\mu}}{[k^{2} -
  (x+y)^{2}m^{2} - (1-x-y)M^{2}]^{3}},
\end{equation}
where
\begin{eqnarray}
\tilde{N}_{1}^{\mu} & = & \gamma^{\mu}\left[m^{2}(4-4(x+y)-(x+y)^{2})
  + k^{2}\right] - 2k^{\mu} k\!\!\!\slash + \frac{m}{2}
  [\gamma^{\mu},q\!\!\!\slash] (2 (x+y) -
 (x+y)^{2}).\label{N1tilde}
\end{eqnarray}
Here, in addition to the fermion equations (as well as setting $q^{2}=0$ in
the denominator), we have used Gordon decomposition,
\begin{equation}
(p'^{\mu} + p^{\mu}) = 2m \gamma^{\mu} + \frac{1}{2}
[\gamma^{\mu},q\!\!\!\slash].
\end{equation}
We note here that because of various identities involving the gamma
matrices in $2+1$ dimensions, the Gordon decomposition can be written
in alternate equivalent ways, but we would continue to use it in the
conventional form given above.

It is only the last term in (\ref{N1tilde}) 
that leads to the magnetic moment. The momentum as well as the Feynman
parameter integrals can be evaluated for this term in a
straightforward manner to give
\begin{equation}
\Gamma_{(\eta)}^{\mu\, ({\rm mag})} (q^{2}=0)
  =  \frac{e}{2m} \sigma^{\mu\nu} q_{\nu}\,F_{2}^{(\eta)}
 (q^{2}=0),
\end{equation}
where
\begin{equation}
F_{2}^{(\eta)} (q^{2}=0) = F_{2}^{(\eta)} (0) = - \frac{e^{2}}{8\pi
  m}\left[3(1-\kappa) - (2- \frac{3\kappa^{2}}{2}) \ln \frac{2 +
  \kappa}{\kappa}\right].\label{eta} 
\end{equation}
This shows that when $\kappa\rightarrow 0$ (namely, $M\rightarrow 0$),
there is a divergence in the magnetic moment in the contribution
coming from  the
$\eta_{\rho\lambda}$ term in the propagator (\ref{equivprop}). This
is an  infrared divergence. In fact,
we note here that when $\kappa = 0$, the contribution to the form
factor  coming from 
$\eta_{\rho\lambda}$ at a general momentum transfer has the form
\begin{equation}
F_{2 (\kappa = 0)}^{(\eta)} (q^{2}) = \frac{e^{2}m}{2\pi
  (q^{2}-4m^{2})}\left(1 +
\frac{m}{\sqrt{q^{2}}}\,\ln \frac{2m + \sqrt{q^{2}}}{2m -
  \sqrt{q^{2}}}\right) - \frac{e^{2}m}{2\pi (q^{2}-4m^{2})}
\int_{0}^{1} \frac{\mathrm{d}x}{x}.
\end{equation}
Namely, when $\kappa = 0$, the form factor is not even defined for any
value of the momentum transfer. Furthermore, when $\kappa\rightarrow
\infty$, we see from (\ref{eta}) that
\begin{equation}
F_{2}^{(\eta)} (0) \rightarrow - \frac{e^{2}}{8\pi m}\left(-
\frac{2}{\kappa^{2}} + O (\frac{1}{\kappa^{3}})\right)\rightarrow 0,
\end{equation}
as it should, since in the limit $\kappa\rightarrow\infty$
($M\rightarrow\infty$), the theory corresponds to a pure Chern-Simons
theory interacting with fermions and the photon propagator does not
have an $\eta_{\mu\nu}$ term in this case.

To obtain the contribution from the $\epsilon_{\rho\lambda\sigma}$
term in the propagator in (\ref{equivprop}), we note that we can simplify the
Dirac algebra using (\ref{gammaidentity}) to obtain the numerator in
(\ref{vertex}) to be 
(just the tensor structure without factors such as $i,M$)
\begin{equation}
N_{2}^{\mu} = N_{2}^{(1)\,\mu} + N_{2}^{(2)\,\mu}\, ,
\end{equation}
where
\begin{eqnarray}
N_{2}^{(1)\, \mu} & = & k^{2}\left(4m \gamma^{\mu} - 2k^{\mu} + 2
\left[\gamma^{\mu},q\!\!\!\slash\right]\right) \,
 ,\nonumber\\
\noalign{\vskip 4pt}%
N_{2}^{(2)\,\mu} & = &  \left[-4m k\cdot
  (p'+p) \gamma^{\mu} + 4k\cdot p' k^{\mu} + 2 k\cdot q k\!\!\!\slash
  \gamma^{\mu} + 4k\cdot p q\!\!\!\slash \gamma^{\mu}\right.\nonumber\\
\noalign{\vskip 4pt}%
 &  & \left. \quad - 4m
q^{\mu}k\!\!\!\slash + 2q\!\!\!\slash \gamma^{\mu}q\!\!\!\slash
k\!\!\!\slash - 4 k^{\mu} k\!\!\!\slash q\!\!\!\slash + 4m (p'^{\mu} +
p^{\mu}) k\!\!\!\slash - 8m k^{\mu} k\!\!\!\slash\right]\, .\label{N2}
\end{eqnarray}
The denominators can now be combined and integration variables shifted
to yield
\begin{eqnarray}
\Gamma_{(\epsilon)}^{\mu} & = & \pm 2e^{3}M \int_{0}^{1}
\mathrm{d}x \int_{0}^{1-x}
\mathrm{d}y\,\mathrm{d}K\left[\frac{\tilde{N}_{2}^{(1)
    \mu}}{(k^{2} - (x+y)^{2}m^{2} - (1-x-y)M^{2})^{3}}\right.\nonumber\\
 &  & + \frac{1}{M^{2}}\left.\, \tilde{N}_{2}^{(2)
  \mu}\left(\frac{1}{(k^{2} - (x+y)^{2}m^{2} - (1-x-y)M^{2})^{3}} -
\frac{1}{(k^{2} - (x+y)^{2}m^{2})^{3}}\right)\right],
\end{eqnarray}
where the terms that can contribute to the magnetic moment in
$\tilde{N}_{2}^{(1) \mu}, \tilde{N}_{2}^{(2) \mu}$ have the forms
\begin{eqnarray}
\tilde{N}_{2}^{(1) \mu\,({\rm mag})} & = &  \frac{1}{2} (4-(x+y))
[\gamma^{\mu},q\!\!\!\slash],\nonumber\\
\tilde{N}_{2}^{(2) \mu\,({\rm mag})} & = & -m^{2} (x+y)^{2}
      [\gamma^{\mu}, q\!\!\!\slash] + 4 k\cdot p' k^{\mu} + 2k\cdot q
      k\!\!\!\slash \gamma^{\mu} - 4k\!\!\!\slash q\!\!\!\slash
      k^{\mu}.\label{N2tilde}
\end{eqnarray}
In deriving these, we have used the equations for the fermions, Gordon
decomposition and we have set $q^{2}=0$ in the denominator. 

With (\ref{N2tilde}), the momentum as well as the Feynman integrals can be
evaluated and we obtain
\begin{eqnarray}
\Gamma_{(\epsilon, N_{2}^{(1)})}^{\mu\, ({\rm mag})} (q^{2}=0)
 & = & \pm \frac{e}{2m}\,\sigma^{\mu\nu} q_{\nu}
 \left(-\frac{e^{2}}{8\pi m}\right)\left\{\frac{8+2\kappa}{2+\kappa}
 - \kappa \ln \frac{2+\kappa}{\kappa}\right\},\nonumber\\
\Gamma_{(\epsilon, N_{2}^{(2)})}^{\mu\, ({\rm mag})} (q^{2}=0) 
 & = & \pm
  \frac{e}{2m}\,\sigma^{\mu\nu} q_{\nu}\left(\frac{e^{2}}{4\pi
 m}\right)\left\{- \frac{\kappa}{2+\kappa}
  + \frac{\kappa}{2} \ln \frac{2+\kappa}{\kappa}\right\}\, .\label{N22}
\end{eqnarray}

Adding the two contributions in (\ref{N22}), we determine the total
contribution to the magnetic moment coming from the
$\epsilon_{\rho\lambda\sigma}$ term in the propagator
\begin{equation}
\Gamma_{(\epsilon)}^{\mu\, ({\rm mag})} (q^{2}=0) = \pm \frac{e}{2m}\,
\sigma^{\mu\nu} q_{\nu} \left(\frac{e^{2}}{4\pi m}\right)
\left\{-2 + \kappa \ln \frac{2+\kappa}{\kappa}\right\}\, .\label{epsilon}
\end{equation}
We note that this is well behaved at $\kappa = 0$ and, in fact, gives
a finite result which signals once again that the integrals must be infrared
divergent (since the coefficient multiplying the integral has a factor
of $M$). Furthermore, as $\kappa\rightarrow \infty$,
\begin{equation}
\left(-2 + \kappa \ln \frac{2+\kappa}{\kappa}\right) \rightarrow
-\frac{2}{\kappa} + O (\frac{1}{\kappa^{2}})\rightarrow 0\, .
\end{equation}
Namely, the contribution from the $\eta_{\mu\nu}$ term vanishes much
faster than that from the $\epsilon_{\mu\nu\lambda}$ term in this
limit. 

Finally, adding (\ref{eta}) and (\ref{epsilon}), we obtain
\begin{eqnarray}
F_{2}^{(+)} (0) & = &  
\left(\frac{e^{2}}{8\pi m}\right) \left\{-(7-3\kappa) + (2+2\kappa -
  \frac{3\kappa^{2}}{2}) \ln \frac{2+\kappa}{\kappa}\right\}\,
,\nonumber\\
\noalign{\vskip 4pt}%
F_{2}^{(-)} (0) & = & 
\left(\frac{e^{2}}{8\pi m}\right) \left\{(1+3\kappa) + (2-2\kappa -
  \frac{3\kappa^{2}}{2}) \ln \frac{2+\kappa}{\kappa}\right\}\,
.\label{zeroT} 
\end{eqnarray}
Here ``$\pm$'' correspond to the theory with the two signs for the
Chern-Simons term. 
\begin{figure}[h]
\begin{center}
\includegraphics{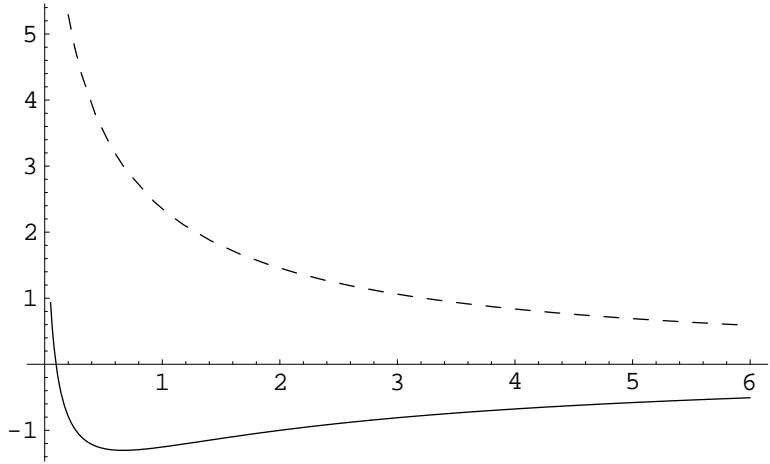}
\caption{The solid and the dashed lines represent respectively
  $F_{2}^{(+)}(0)$ and $F_{2}^{(-)}(0)$, in units of $\frac{e^{2}}{8\pi m}$,
  as a function of $\kappa$.}
\end{center}
\end{figure}

Let us note from (\ref{zeroT}) that $F_{2}^{(+)}
(0)$ 
has a minimum at $\kappa = \frac{2}{3}$, where it has the value
\begin{equation}
F_{2}^{({\rm min})\,(+)} (0) = (-5 + \frac{8}{3} \ln
    4)\left(\frac{e^{2}}{8\pi m}\right) < 0\, .
\end{equation}
On the other hand, $F_{2}^{(-)}$ smoothly vanishes at infinity and 
does not have a nontrivial extremum. Both the form factors diverge at
$\kappa = 0$ signalling an infrared divergence. Asymptotically, as
$\kappa$ becomes large, they behave as
\begin{equation}
F_{2}^{(\pm)} (0) \rightarrow \frac{e^{2}}{2\pi m} \left(\mp
  \frac{1}{\kappa}\right).
\end{equation}
This indeed agrees with the large $M$ result in \cite{kogan}. In fact,
recalling that the pure Chern-Simons theory can be obtained from the
Maxwell-Chern-Simons theory in the limit $e,M\rightarrow \infty$, such
that $\frac{e^{2}}{M}$ is fixed, we obtain the well known result
\cite{kogan,girotti} that
\begin{equation}
F_{2}^{({\rm CS})\,(\pm)} (0) = \mp \frac{e^{2}}{2\pi M}\,
.\label{pureCS} 
\end{equation}
The zero temperature propagator for the photon in the pure Chern-Simons
theory in the covariant gauge has the form
\begin{equation}
i D_{\mu\nu} (p) = - \frac{i}{p^{2}} \left(\xi
\frac{p_{\mu}p_{\nu}}{p^{2}} \mp \frac{i}{M} \epsilon_{\mu\nu\lambda}
p^{\lambda}\right)\, .\label{CS}
\end{equation}
As we have argued earlier, the $p_{\mu}p_{\nu}$ term does not contribute
to the magnetic moment and the $\frac{1}{M}$ dependence in
(\ref{pureCS}) is then easily understood as arising simply from the
coefficient of the $\epsilon_{\mu\nu\lambda}$ term in the propagator
in (\ref{CS}). The anomalous magnetic moment in the pure Chern-Simons
case has been argued  in \cite{kogan} to result from an ``induced spin''.

This zero temperature analysis clearly shows that in the
Maxwell-Chern-Simons theory, the anomalous
magnetic moment (\ref{zeroT}) has an infrared divergence for
$M=0$ (namely, in the usual QED without a Chern-Simons term). This
suggests,
therefore, that the Chern-Simons term is a necessity in such theories
to have  well defined physical quantities such as the anomalous
magnetic moment. Furthermore, we note that the magnitude of the
anomalous magnetic moment in such a theory can become arbitrarily
large depending on
the value of $M$, unlike in $3+1$ dimensional QED.

\section{Finite Temperature}

The finite temperature analysis of the anomalous magnetic moment is
quite a bit more involved. We have carried out our calculations both
in the imaginary time as well as in the real time formalisms. However,
we will 
describe the details only in the real time formalism for
simplicity \cite{das}. We note that at finite temperature, isolating
the magnetic
moment contribution from the vertex needs care since at finite
temperature, the vertex can have other tensor structures \cite{yee},
for example, of the general form (compare with (\ref{zeroTstructure}))
\begin{equation}
\Gamma^{\mu} (p,p') = e\gamma^{\mu}\,F_{1} (q^{2}) +
\frac{e}{2m}\,\sigma^{\mu\nu} q_{\nu}\,F_{2} (q^{2}) + u^{\mu} (u\cdot
\gamma)\,F_{3} (q^{2})\, ,
\end{equation}
where $u^{\mu}$ represents the velocity of the heat bath. We will work
with the heat bath at rest so that the relevant components of the
gauge and fermion propagators will have the forms
\begin{eqnarray}
iD_{\mu\nu}^{(\beta)\,({\rm eff})} (p) & = & - \left(\eta_{\mu\nu} \pm
iM\epsilon_{\mu\nu\lambda}
\frac{p^{\lambda}}{p^{2}}\right)\left[\frac{i}{p^{2}-M^{2}} + 2\pi
  n_{B} (|p^{0}|) \delta (p^{2}-M^{2})\right]\nonumber\\
\noalign{\vskip 4pt}%
iS^{(\beta)} (p) & = & (p\!\!\!\slash + m)\left[\frac{i}{p^{2}-m^{2}}
  - 2\pi n_{F} (|p^{0}|) \delta (p^{2}-m^{2})\right]\,
,\label{Tpropagators}
\end{eqnarray}
where
\begin{equation}
n_{B} (x) = \frac{1}{e^{\beta x} -1},\qquad n_{F} (x) =
\frac{1}{e^{\beta x} + 1}\, ,
\end{equation}
represent the Bose-Einstein and the Fermi-Dirac distribution functions
respectively with $\beta$ denoting the inverse temperature in units of
the Boltzmann constant.

In studying the question of thermal corrections to the anomalous
magnetic moment, we are interested in the scattering of
low momentum, on-shell electrons from a static magnetic field. We note
that in the real time formalism, since the propagators
separate into a zero temperature part and a finite temperature part,
the finite temperature correction to the vertex in (\ref{vertex}) can
be identified in a simple manner. Furthermore, since the thermal integrals
cannot be evaluated in closed form, we will be considering the high
temperature limit 
\begin{equation}
m, M \ll \frac{1}{\beta}\, .\label{highT}
\end{equation}
In the high temperature limit, terms with a single thermal
propagator (namely, with only one distribution function) give the
dominant contribution and the results from the imaginary time and the
real time formalisms coincide \cite{brandt2}. (In this limit, both the
formalisms give a retarded contribution.) 

With this, we are ready to evaluate the thermal corrections to the
anomalous magnetic moment. We note that the numerator in
(\ref{vertex}) continues to be the same at finite temperature so that
we can take over the Dirac algebra from the zero temperature
calculations. The delta function integral (in the thermal part of the
propagator) can, of course, be done trivially and the remaining
integrals can be evaluated, in the low momentum approximation, with
the series representations
\begin{equation}
\frac{1}{e^{x}-1} = \frac{1}{x} - \frac{1}{2} + 2 \sum_{n=1}^{\infty}
\frac{x}{x^{2}+ (2\pi n)^{2}},\quad \frac{1}{e^{x}+1} = \frac{1}{2} -
\sum_{n=-\infty}^{\infty} \frac{x}{x^{2} + ((2n + 1)\pi)^{2}}\, .
\end{equation}
It follows from this that in the high temperature limit, the
contribution coming from the bosonic distribution function gives the
leading term. Therefore, we will list below only the relevant integrals
involving the bosonic distribution function in the high temperature
limit and low momentum expansion. Let us define
\begin{equation}
J = \delta (k^{2}-M^{2})\,n_{B} (|k^{0}|)\,\frac{1}{(k^{2}-2k\cdot
  p)(k^{2}-2k\cdot p')}\, .\label{J}
\end{equation}
Then, in the high temperature limit and in the low momentum expansion,
we have
\begin{eqnarray}
I & = & \int \mathrm{d} K\,J \approx \frac{1}{2 (2\pi)^{2} \beta m^{4}} 
\frac{1}{\kappa^{2}}\left[\frac{2}{4-\kappa^{2}} +
  \frac{1}{\kappa^{2}}\ln \frac{4-\kappa^{2}}{4}\right] + O
(\beta^{0}),\nonumber\\
\noalign{\vskip 4pt}%
 I^{0} & = & \int \mathrm{d} K\,k^{0} J \approx \frac{1}{4 (2\pi)^{2}
   \beta m^{3}}\,\frac{1}{4-\kappa^{2}} + O
 (\beta^{0}),\nonumber\\
\noalign{\vskip 4pt}%
I^{i} & = & \int \mathrm{d} K\,k^{i} J \approx \frac{(p+p')^{i}}{8
  (2\pi)^{2} \beta m^{4}}\left[\frac{2}{4-\kappa^{2}} +
  \frac{1}{\kappa^{2}} + \frac{4}{\kappa^{4}} \ln
  \frac{4-\kappa^{2}}{4}\right] + O (\beta^{0}),\nonumber\\
\noalign{\vskip 4pt}%
I^{00} & = & \int \mathrm{d} K\,k^{0}k^{0} J \approx - \frac{1}{4
  (2\pi)^{2} \beta m^{2}}\,\ln \beta M + O(\beta^{-1}),\nonumber\\
\noalign{\vskip 4pt}%
I^{0i} & = & \int \mathrm{d} K\,k^{0}k^{i} J \approx -
\frac{(p+p')^{i}}{8 (2\pi)^{2} \beta m^{3}}\, \ln \beta M + O
(\beta^{-1}),\nonumber\\
\noalign{\vskip 4pt}%
I^{ij} & = & \int \mathrm{d} K\, k^{i}k^{j} J \approx
-\frac{\delta^{ij}}{8 (2\pi)^{2} \beta m^{2}}\, \ln \beta M +
O(\beta^{-1})\, .\label{integrals}
\end{eqnarray}

There are several things to note from the structure of the integrals
in (\ref{integrals}). First, clearly the dominant temperature
dependence comes from integrals with $k_{\mu}k_{\nu}$ in the
numerator. We note that the integrals (in particular, the subleading
terms)  are, in fact, divergent when $\kappa =
0,2$. The divergence at $\kappa = 0$ is easily understood. It is the
infrared divergence we have already encountered at zero temperature,
but is much more severe at finite temperature. The divergence at
$\kappa = 2$ ($M=2m$), however, is new and, therefore, let us discuss
this briefly. At first sight, this seems like a threshhold singularity
for a particle of mass $M$ to decay to two particles each of mass
$m$. To understand this better, let us examine the first integral in
(\ref{integrals}). With the definition in (\ref{J}), we note that we
can integrate out the delta function and write the leading term in the
low momentum expansion as
\begin{eqnarray}
I & = & \frac{1}{2 (2\pi)^{3}} \int \mathrm{d}^{2}
k\,\frac{n_{B}(\omega)}{\omega} \Big[\frac{1}{(M^{2} - 2\omega p^{0} +
    2{\bf k}\cdot {\bf p})(M^{2} - 2\omega p'^{0} + 2 {\bf k}\cdot
    {\bf p}')}\nonumber\\
\noalign{\vskip 4pt}%
 &  & \qquad\qquad + (p^{0},p'^{0}\rightarrow -p^{0},
-p'^{0})\Big]\nonumber\\
\noalign{\vskip 4pt}%
 & \approx & \frac{1}{2 (2\pi)^{2}} \int_{M}^{\infty}
\frac{\mathrm{d}x}{e^{\beta x} - 1} \left[\frac{1}{(M^{2} - 2mx)^{2}}
  + \frac{1}{(M^{2} + 2mx)^{2}}\right]\, ,\label{analyze}
\end{eqnarray}
where $\omega = \sqrt{{\bf k}^{2}+M^{2}}$. It is clear from the first
line in (\ref{analyze})  that the integrand has a singularity at
\begin{equation}
\omega_{*} = \frac{p^{0}M^{2} \pm pM \cos\theta \sqrt{M^{2} - 4
    (m^{2}+p^{2}\sin^{2}\theta)}}{2 (m^{2} + p^{2} \sin^{2} \theta)}\,
    ,
\end{equation}
where $p = |{\bf p}|$. Namely, there is a branch cut for
\begin{equation}
M^{2} \geq 4 (m^{2} + p^{2})\, .
\end{equation}
In the leading term in the low momentum expansion, on the other hand,
we  see from the
last line in (\ref{analyze}) that the singularity is a double pole at
$x = \frac{M^{2}}{2m}$ which becomes a genuine end point singularity
\cite{eden} for
$M=2m$. For this value of $M$, the low momentum expansion breaks down and,
consequently, we will avoid such a value in our calculation. (We also
want to emphasize that the arguments of the logarithms should be
considered with $m\rightarrow m \mp i\epsilon$ where $i\epsilon$ comes
from the Feynman prescription that we have been suppressing.)

We can now obtain the leading order thermal corrections to the
anomalous magnetic moment which, as we have noted, come from quadratic
powers of $k$ in the numerator. Thus, looking at (\ref{N1}), we note
that the leading contribution coming from the $\eta_{\mu\nu}$ terms in the
photon propagator will have a form
\begin{equation}
\Gamma_{(\eta)}^{i\,({\rm mag})} = - 4i\pi e^{3} \left(\gamma^{0}
I^{0i} - \gamma^{j} I^{ij}\right)\, .
\end{equation}
The magnetic form factor can now be read out using (\ref{integrals}),
fermion equations as well as Gordon decomposition, to give
\begin{equation}
F_{2}^{(\beta)\,(\eta)} (q^2 = 0) = \frac{e^{2}}{4\pi
    m}\,\frac{1}{\beta m}\,\ln \beta M + O (\beta^{-1})\, .
\end{equation}
The leading contributions coming from the
$\epsilon_{\mu\nu\lambda}$ terms in the propagator can, similarly be
obtained in a straightforward manner. First, we note that the
$\epsilon_{\mu\nu\lambda}$ term has an extra factor of $k^{2}$ in the
denominator which will cancel with the overall $k^{2}$ in
$N_{2}^{(1)\,\mu}$ in (\ref{N2}).  As a result, 
we recognize that there will be no leading order contribution coming
from this term (it has no quadratic terms after canceling the $k^{2}$ in
the denominator). For $N_{2}^{(2)\,\mu}$, we note that the extra
$k^{2}$ in the denominator simply becomes $M^{2}$ because of the delta
function. Consequently, it can have a leading contribution. The
quadratic terms in $k$ in the numerator lead to
\begin{equation}
\Gamma_{(\epsilon, N_{2}^{(2)})}^{i\,({\rm mag})} = \mp \frac{2i\pi}{M}
\left(4(p'^{0} - \gamma^{0}q\!\!\!\slash - 2m)I^{0i} -
2q^{j}\gamma^{0}\gamma^{i}I^{0j} - 4(p'^{j} - \gamma^{j} q\!\!\!\slash
- 2m \gamma^{j}) I^{ij} + 2 q^{j}\gamma^{k}\gamma^{i} I^{jk}\right)\,
.\label{N2T}
\end{equation}
It is straightforward to check using (\ref{integrals}) (and Gordon
decomposition) that the leading terms cancel out in (\ref{N2T}) so
that  
\begin{equation}
F_{2}^{(\beta)\,(\epsilon)} (q^{2} = 0) = O (\beta^{-1})\, .
\end{equation}
As a result, the leading order thermal correction to the magnetic
moment in the scattering of low momentum electrons from a
static magnetic field comes from the $\eta_{\mu\nu}$ terms in the
photon propagator and has the form
\begin{equation}
F_{2}^{(\beta)} (q^{2}=0) = \frac{e^{2}}{4\pi m}\, \frac{1}{\beta
  m}\,\ln \beta M\, .\label{Tresult}
\end{equation}
This can be compared with the leading thermal behavior of the anomalous
magnetic moment for the electron in $3+1$ dimensional QED which has a
$\frac{1}{\beta^{2}}$ dependence \cite{yee,barton}. The leading order
thermal behavior
in (\ref{Tresult}) is independent of the sign of the Chern-Simons
coefficient since the $\epsilon_{\mu\nu\lambda}$ term gives a
subleading contribution. Equation (\ref{Tresult})
shows that the leading term in the anomalous magnetic moment is
infrared divergent as are the subleading terms (from the structures in
(\ref{integrals})).  

Since our result was derived in the high temperature limit using
(\ref{highT}), we cannot obtain the magnetic moment for the pure CS
theory from (\ref{Tresult}). But calculating the contribution in the
pure CS theory is relatively simple. We note from (\ref{CS}) that, at
finite temperature, the photon propagator has the form
\begin{equation}
iD_{\mu\nu}^{(\beta)} (p) = - \left(\xi \frac{p_{\mu}p_{\nu}}{p^{2}}
\mp \frac{i}{M} \epsilon_{\mu\nu\lambda}
p^{\lambda}\right)\left(\frac{i}{p^{2}} + 2\pi n_{B} (|p^{0}|) \delta
(p^{2})\right)\, .\label{CST}
\end{equation}
As we have argued, it is only the $\epsilon_{\mu\nu\lambda}$ term in
(\ref{CST})  that will contribute to
the magnetic moment. The numerator of the integrand still has the same
structure as in (\ref{N2}). It is easy to see that for the bosonic
distribution function $N_{2}^{(1)\,\mu}$ will give vanishing
contribution because of the delta function in (\ref{CST}). The
contribution coming from $N_{2}^{(2)\,\mu}$ has the form
\begin{equation}
\Gamma_{N_{2}^{(2)}}^{i\,({\rm mag})} = \mp \frac{4i\pi e^{3}}{m^{2}}
\left(-(2p'^{i}+p^{i}) - \frac{1}{2} q\!\!\!\slash \gamma^{i} +
\gamma^{i} q\!\!\!\slash\right) \int \mathrm{d}K\,\delta (k^{2})
n_{B}(|k^{0}|)\, .\label{bosonic}
\end{equation}
Using Gordon decomposition, it follows that this gives a vanishing
contribution to the magnetic moment. 

As a result, we have to look at
the contributions coming from the fermion distribution function. Once
again, it can be shown that the integral involving $N_{2}^{(1)\,\mu}$
vanishes identically. The contribution coming from $N_{2}^{(2)\,\mu}$,
in the low momentum expansion, 
can be derived in a straightforward manner and yields
\begin{equation}
\Gamma_{N_{2}^{(2)}}^{i\, ({\rm mag})} = \pm \frac{i\pi e^{3}}{M} \int
\mathrm{d}K\,\delta (k^{2}-m^{2}) n_{F} (|k^{0}|)
\left(\frac{4(q\!\!\!\slash \gamma^{i} + 2p^{i})}{m^{2} + mk^{0}}
+\frac{(2\gamma^{i}q\!\!\!\slash - q\!\!\!\slash \gamma^{i} -
(4p'^{i}+2p^{i})) \mathbf{k}^{2}}{(m^{2}+mk^{0})^{2}}\right)\,
.\label{fermionic}
\end{equation}
Upon using Gordon decomposition, this gives a vanishing contribution
to the magnetic moment. We note that the coefficients for the magnetic
moment in
(\ref{bosonic}) and (\ref{fermionic}) vanish even before doing the
integral and, therefore, this result holds for any nonzero
temperature. This will also hold
even if the integrals have divergences because 
they can always be regularized (The infrared divergence can be easily
regularized, for example, by including a regulator mass in the delta
function.), with the  coefficients leading to a vanishing
result. Therefore, we conclude that there is no finite
temperature correction to the anomalous magnetic moment in the pure CS
theory. At first sight, this is unexpected, but it can be intuitively
understood as follows. As has been argued \cite{kogan}, the zero temperature
anomalous magnetic moment in the pure CS theory can be understood as
an induced spin
effect. However, we do not expect temperature to modify the spin
behavior of a particle and, therefore, the vanishing of the thermal
correction to the anomaluos magnetic moment in the pure CS theory does
make sense. However, at present we have no other fundamental argument
for why it
should vanish. It remains an interesting question to see if it will
continue to vanish at higher loops. (We note that while the one loop
temperature dependence of the CS term \cite{das1} will lead to a temperature
dependence at higher loops, there will be several other sources of
temperature dependence and it is not clear {\it a priori} whether they
will all cancel.)

\section{Conclusion}

We have systematically studied the anomalous magnetic moment of the
electron  in the $2+1$ dimensional Chern-Simons theory with or without
a Maxwell term, both at zero and at finite temperature. At zero
temperature, we find that the anomalous magnetic moment in the
Maxwell-Chern-Simons theory is divergent unless there is a tree level
Chern-Simons coefficient. This suggests strongly that a Chern-Simons
term is necessary in such theories. In the limit of large $M$, we
recover the earlier known results of pure CS theory
\cite{kogan,girotti}.  At finite
temperature, we find that the anomalous magnetic moment in the MCS
theory behaves, at high
temperature, as $\frac{1}{\beta} \ln \beta M$. The results
show a strong infrared divergence for $M\rightarrow 0$. Just for
completeness, we note here that the high temperature behavior of the
Chern-Simons term at one loop 
in such theories  goes as $\beta$ in the static limit \cite{das1}
(or $\beta \ln \beta m$ in the long wave limit \cite{frenkel}). The pure
Chern-Simons theory is even more interesting in that there is no
thermal correction to the anomalous magnetic moment at one loop. We
give a plausibility argument for why this is natural, but a better
understanding of this interesting question remains open.

\vskip .5cm

We would like to thank Professors G. Dunne, J. Frenkel and J. C. Taylor
for comments and suggestions. This work was supported in part by US
DOE Grant number DE-FG 02-91ER40685.

\end{document}